# On the role of the Michelson-Morley experiment: Einstein in Chicago

*Jeroen van Dongen*

Institute for History and Foundations of Science & Descartes Centre, Utrecht University & Einstein Papers Project, Caltech

Abstract:
*This article discusses new material, published in Volume 12 of the Collected Papers of Albert Einstein, that addresses Einstein's knowledge of the Michelson-Morley experiment prior to 1905: in a lecture in Chicago in 1921, Einstein referred to the experiment, mentioned when he came upon it, and hinted at its influence. Arguments are presented to explain the contrast with Einstein's later pronouncements on the role of the experiment.*

In the history of the special theory of relativity, one of the most debated topics has been the precise role played by the Michelson-Morley experiment in Einstein's creation of the theory. In the early twentieth century, it was established lore that the Michelson-Morley experiment played a central and constitutive role—that, in fact, the special theory was essentially formed through a "generalization from Michelson's experiment,"[1] in the words of Robert A. Millikan, once a close junior colleague of Michelson at the University of Chicago.

Yet, Einstein did not explicitly mention the experiment in his 1905 publication,[2] and this circumstance raises questions regarding the accurateness of the old lore. Einstein did of course take up the Michelson-Morley experiment on numerous occasions in later publications. He cited it for the first time in a review article of 1907,[3] and, as another example, in 1910, in another review, he discussed the experiment to illustrate the unnaturalness of Lorentz's length contraction hypothesis.[4]

Most of these discussions were related to the question of the evidential support for special relativity.[5] There are far fewer sources that illuminate the concrete role of the experiment in Einstein's thought leading up to 1905, and those that are extant are not always unambiguous. We will discuss a new addition to these documents here.

In 1969, Gerald Holton made the first detailed study of historical sources then available and argued that the Michelson-Morley experiment had played a limited and indirect part: according to Einstein's recollections later in life, the role it had was at best to allay any doubts that he may have had about the general validity of the principle of

---

[1] (Millikan 1949), on p. 343.
[2] (Einstein 1905).
[3] (Einstein 1907), p. 412.
[4] (Einstein 1910), pp. 10-16.
[5] See also the discussion in (Stachel 1982/2002).

relativity, or indeed to emphasize the inadequacy of Lorentz's theory. Einstein, himself, for instance, claimed in 1942 that he had already become "pretty much convinced of the validity of the [relativity] principle before I did know this experiment and its results."[6] Another fairly typical statement by the older Einstein of his recollection is:

In my own development Michelson's result has not had a considerable influence. I even do not remember if I knew of it at all when I wrote my first paper on the subject (1905). The explanation is that I was, for general reasons, firmly convinced that there does not exist absolute motion and my problem was only how this could be reconciled with our knowledge of electro-dynamics. One can therefore understand why in my personal struggle Michelson's experiment played no role or at least no decisive role.[7]

The most important resource on Einstein's early years that has become available since Holton's groundbreaking work is the 42 letters that Einstein exchanged with his fiancée Mileva Marić between 1898 and 1902.[8] The letters provide indirect evidence that Einstein did at least know of the experiment by 1899. Yet, they do not mention the experiment explicitly. In particular, in a letter from 1899 Einstein reported that he read Wilhelm Wien's review on translatory motion of the ether. Wien's article contained a description of 13 experiments on the problem, including a two paragraph outline of the Michelson-Morley experiment.[9] The letters also indicate that Einstein had thought out some ether drift experiments of his own, without however revealing any substantial details. John Stachel, editor of the first volume of the Collected Papers of Albert Einstein, agreed with Holton's observation that the Michelson-Morley experiment had played no significant role, but also argued that ether drift experiments were an important strand in the complex of ideas that eventually led to the special theory.[10]

Einstein traveled to Japan in 1922, and at Kyoto University gave a lecture entitled "How I created the theory of relativity." To date, this lecture has been the most detailed source of information concerning Einstein's (non-)involvement with the Michelson-Morley experiment, but there is no universal agreement about its precise content. In a recent rendition,[11] Einstein is quoted as having said that he initially did not doubt the movement of the earth relative to the ether, and that he thought up an experiment to test this assertion. In the arrangement, light from a single source was to be split into two light beams moving parallel and in opposite direction to the earth's motion; the presumed difference in the energy of the two beams was to be measured by thermocouples. Einstein added: "This idea was of the same sort as that of Michelson's experiment, but I did not know this experiment very well then." Yet, his ignorance would not last long: "While I had these ideas in mind as a student, I came to know the strange result of the Michelson experiment. Then I came to realize intuitively that, if we admit this as a fact, it must be our mistake to think of the movement of the earth against the ether. That was the first route that led me to what we now call the principle of special relativity."[12]

---

[6] Einstein to Michelson's biographer, Bernard Jaffe, on 17 March 1942; as in (Holton 1969/1995), p. 340.
[7] Einstein to F.G. Davenport, 9 February 1954, EA (i.e. Einstein Archive, Jerusalem) 17 199. See also (Holton 1969/1995), pp. 343-344.
[8] They can be found in CPAE 1; see also (Renn and Schulmann 1992).
[9] (Wien 1898), the Michelson-Morley experiment is discussed on its pp. xv-xvi.
[10] (Stachel 1987/2002), see in particular p. 175.
[11] See (Abiko 2000); a translation of Einstein's lecture is on its pages 13-17.

This passage suggests that the Michelson-Morley experiment did, after all, influence Einstein in a direct way, and was relevant in the construction of the theory. Nevertheless, considerable debate has persisted, focusing on the translation of the Kyoto address. Notes of the lecture were taken in Japanese by Einstein's interpreter, Jun Ishiwara; two English translations appeared in 1979 and 1982.[13] These translations indeed implied a direct role for the Michelson-Morley experiment, but they have been criticized by Seiya Abiko and Ryoichi Itagaki. Regrettably, these critics are however not agreed on what a proper translation should be. According to Itagaki, the above passage, taken from Abiko (p. 13), should read: "But when, still as a student, I had these thoughts in my mind, if I had known the strange result of this Michelson's experiment and I had acknowledged it as a fact, I probably would have come to realize it intuitively as our mistake to think of the motion of the Earth against the ether."[14]

For scholars who do not master the Japanese language, and given that Einstein actually delivered his lectures in German, it is thus difficult now to know Einstein's precise words in Kyoto on the Michelson-Morley experiment.[15] Fortunately, recently uncovered documents show that, a year earlier, Einstein addressed the issue on a similar occasion in Chicago.

During his first trip to the United States, Einstein visited Chicago in early May of 1921. The main purpose of his visit was to raise funds for the creation of a Hebrew University in Jerusalem, one of the main causes of the Zionist movement that had enlisted Einstein's support.[16] The Zionist leadership had asked him to accompany a delegation and had encouraged him to lecture at American universities,[17] leading Einstein to agree to an engagement at the University of Chicago. Einstein would give three lectures at the university and one at a local school, the Francis W. Parker School.[18]

Einstein hoped that his visit would give him the opportunity to meet local scientists, most prominent among whom were Robert A. Millikan and Albert A. Michelson.[19] Michelson was away, traveling in Europe, but Millikan, who had not yet taken up his position as president of the California Institute of Technology, was present. By one account, Millikan "loved and venerated" Einstein from the moment they met.[20]

---

[12] Einstein, as in Abiko, *ibid.* on p. 13.
[13] Jun Ishiwara, pp. 131-133 in *Ainsutain Koenroku* (Tokyo: Kaizo-sha, 1923); (Ogawa 1979), (Einstein 1922/1982).
[14] (Itagaki 1999).
[15] The future Volume 13 of the Collected Papers of Albert Einstein is to provide another, authoritative translation of the Kyoto lecture.
[16] See the Introduction of CPAE 12, section I, and (Rosenkranz 2009).
[17] See Chaim Weizmann to Einstein, 9 March 1921 (CPAE 12, Doc. 91).
[18] For a republication of press reports, see (Illy 2006), in particular its chapter "Baffled in Chicago", pp. 146-159.
[19] Einstein to Carl Beck, 8 April 1921 (CPAE 12, Doc. 115); 15 April 1915 (CPAE 12, Doc. 120).
[20] According to Ludwik Silberstein to Einstein, 4 September 1921, (CPAE 12, Doc. 229). On Michelson's absence from Chicago, see (Illy 2006), p. 153; see also Ludwik Silberstein to Einstein, 18 July 1921, (CPAE 12, Doc. 187); on Millikan's presence, see Einstein to Hendrik Antoon Lorentz, 30 June 1921, (CPAE 12, Doc. 163).

The three lectures that Einstein gave at the University of Chicago, presumably on May 3, 4 and 5, resembled the popular lectures that he would give later that month in Princeton. To be precise: the written record of their contents does not closely follow the well-known published versions of the Princeton lectures, but does show a rough similarity with a stenographer's notes of Einstein's two popular lectures in Princeton of May 9 and 10; they are different from his technical talks in Princeton of May 11 through 13. In the case of the lectures at the University of Chicago, a typescript of an auditor's shorthand notes, prepared by a Miss Zimmerman, is all that remains.[21] But Einstein's lectures at the University are actually not of immediate concern here; the talk that he gave at the Francis W. Parker School, however, is.

The record of Einstein's spoken words at the Parker school, likely on May 4, fits on just two pages, which is less than is the case for his lectures at the university.[22] The notes, presumably taken by the same person, were also transcribed by Miss Zimmerman. This typed text may not be a complete record of Einstein's actual words: it is likely that he discussed his route to the general theory of relativity at greater length than the single sentence available: "Then I made the second discovery of my life, that is the discovery of inertia and of gravity."[23]

However, Einstein probably only delivered a very brief talk: at the beginning, he announced that he would only address "a few words" to his audience. This gives greater confidence in the reliability of the recorder's rendition of Einstein's words, despite its brevity. Nevertheless, the somewhat confused nature of the statement on the general theory does emphasize that one should be alert to the possibility that the quality of the rendition of the Parker lecture is not necessarily consistent.

In any case, the notes do contain a full and coherent paragraph on Einstein's early thinking in relation to the special theory of relativity:

As a young man I was interested, as a physicist, in the question what is the nature of light, and, in particular, what is the nature [?] of light with respect to bodies. That is, as a child I was already taught that light is subordinate to the oscillations of the light ether. If that is the case, then one should be able to detect it, and thus I thought about whether it would be possible to perceive through some experiment that the earth moves in the ether. But when I was a student, I saw that experiments of this kind had already been made, in particular by your compatriot, Michelson. He proved that one does not notice anything on earth that it moves, but that everything takes place on earth as if the earth is in a state of rest.[24]

---

[21] These transcripts are contained in the archive of the Wisconsin Historical Society (Madison, WI), Anita McCormick Blaine Papers, Series 1E, Box 237; they are reproduced in CPAE 12, Appendix D; see also this volume's Introduction, section II, for an account of Einstein's visit to Chicago. Note that these documents state that Einstein gave his lectures at the University of Chicago on May 4, 5 and 6; however, the *University of Chicago Magazine*, May edition, and the *Chicago Daily Tribune*, May 2 edition, give as dates May 3, 4 and 5 (see Illy 2006, pp. 146, 152). For the notes of Einstein's popular Princeton lectures, see Appendix C in CPAE 7; an account of his technical lectures is reproduced in Appendix E of CPAE 12, their final published version is (Einstein 1922 [CPAE 7, Doc. 71]).

[22] "Vortrag des Herrn Prof. Dr. Albert Einstein, Chicago, 5. Mai. 1921, Francis Parker School", archive of the Wisconsin Historical Society (Madison, WI), Anita McCormick Blaine Papers, Series 1E, Box 237; see also CPAE 12, Appendix D. An article in the *Chicago Daily Tribune*, "Audience soon sinks in Einstein relativity sea", 4 May 1921, states that the lecture took place on May 4. The transcripts of the three University lectures are resp. 10, 9 and 7 pages long, and they are included together with the lecture at the Parker school in one 28 page long, sequentially paginated document.

[23] "Dann machte ich die zweite Entdeckung meines Lebens, das ist die Entdeckung der Traegheit und der Schwere." See "Vortrag des Herrn … etc."

The detail and coherence of the above passage suggests that it is likely complete and accurate in reproducing Einstein's spoken words. In the next paragraph, Einstein continued with his comment on the general theory given above.

What does the Parker school lecture imply for our understanding of Einstein's relation to the Michelson-Morley experiment, and its influence on the creation of the special theory? Taking the text at face value, there can be no doubt that Einstein knew of the Michelson-Morley experiment prior to 1905. He attributed a significant role to ether drift experiments in general, and singled out the Michelson-Morley experiment for specific mention. It further suggests that Einstein had learned of the experiment *before* becoming convinced of the principle of relativity—contrary to his later recollections.

Einstein's reference to his student days leads one to believe that he may have had in mind his ideas for an ether drift experiment of 1899: his comments perhaps concern the experiment that he briefly mentioned in a letter to Mileva of 10 September 1899—"A good idea occurred to me in Aarau about a way of investigating how the bodies' relative motion with respect to the luminiferous ether affects the velocity of propagation of light in transparent bodies."[25] His next letter to Mileva, dated to 28 September 1899,[26] is the letter that informed her that he had just read Wien's article, i.e. the article containing the Michelson-Morley experiment, among a discussion of other experiments, which confirms Einstein's usage of the plural—"experiments of this kind"—in his lecture in Chicago.

Einstein made no further mention of work on the electrodynamics of moving bodies until 1901.[27] But by then he was no longer a student, leaving the 1899 episode as the most likely candidate on record for a literal reading of his Chicago remarks. Such a literal reading becomes more attractive if one compares the Parker School lecture with Abiko's translation of the Kyoto lecture: the two seem to be in good agreement in suggesting a time frame—his student years—for Einstein's first encounter with the Michelson-Morley experiment. Both texts further suggest a similar role for it: Einstein had formed his own ideas about ether drift experiments, yet these came to naught once he came across other similar experiments and, in particular, the Michelson-Morley experiment. This comparison gives added weight to the above interpretation and enhances confidence in Abiko's translation, particularly given that the two lectures were only about a year apart.[28]

---

[24] "Als junger Mensch hatte mich als Physiker die Frage interessiert, welches das Wesen des Lichtes ist, und im Speziellen, was die Beschaffung des Lichtes zum Koerper ist [sic], naemlich es war so, als Kind schon lehrte man mich, dass das Licht den Schwingungen des Lichtaethers untersteht. Wenn das der Fall ist, so muss man das merken koennen, und so dachte ich darueber nach, ob es moeglich sei, durch irgendwelche Experimente zu merken, dass die Erde im Lichtaether bewegt ist. Aber als ich im Studium war, da sah ich, dass Experimente dieser Art schon gemacht worden waren, insbesondere von Ihrem Landsmann, Michelson. Dieser bewies, dass man auf der Erde nichts davon merke, dass sie sich bewegt, sondern dass auf der Erde alles so vergeht, wie wenn die Erde sich im Zustande der Ruhe befinde." As in "Vortrag des Herrn … etc."

[25] Einstein to Mileva Marić, 10 September 1899 (CPAE 1, Doc. 54).

[26] Einstein to Mileva Marić, 28? September 1899 (CPAE 1, Doc. 57).

[27] See the editorial note "Einstein on the electrodynamics of moving bodies", pp. 223-225 in CPAE 1.

[28] Note however that the description of the experiment in Einstein's letter to Mileva of 10 September 1899 does not in an obvious manner match the experiment outlined in the Kyoto lecture.

There is an evident contradiction between Einstein's description in Chicago of the role played by the Michelson-Morley experiment and his own words of some decades later. How is this contradiction to be understood? Did Einstein let down his guard in Chicago and allow a historical inaccuracy to slip in—perhaps to please Michelson's home crowd? His visit was indeed a great success: soon upon his return to Berlin, Einstein received an informal inquiry as to whether he would be interested joining the University of Chicago physics faculty, an offer which he described as attractive but which he politely declined.[29]

When weighing the different accounts that Einstein gave of his relation to the Michelson-Morley experiment, there are good reasons to accord greater credence to his earlier words, spoken in Chicago and, presumably, Kyoto. One compelling reason is that these fit in well with the contemporary evidence, as argued above. Another, less direct but equally strong reason is that recent scholarship has shown that the later Einstein's recollections of the development of his own research—in particular with regard to the general theory of relativity—were often colored by his subsequent philosophical beliefs and the research program he was pursuing at the time; this would be the highly theoretical unified field theory program, that had virtually no exchange with empirical science.[30] This circumstance coincided with his much discussed pilgrimage from an empiricist philosophical position to that of a "believing rationalist" who seeks the unification of natural forces by exclusively mathematical creativity.[31] In that process, the role of experiment lost prominence in both Einstein's practice in physics, as well as in his philosophical thought.[32]

The difference in emphasis that Einstein accorded the role of the Michelson-Morley experiment in the early nineteen-twenties, compared to his later pronouncements, may naturally be seen as an expression of this broader development in his epistemology. To be sure, in later pronouncements Einstein did attribute a strong role to the Fizeau experiment and the phenomenon of stellar aberration for the construction of the special theory—for instance in his message of 19 December 1952 to, surprisingly, the centenary celebration of Michelson's birth in Cleveland.[33] A constitutive role for the Michelson-Morley experiment had however been central to early empiricist accounts of relativity.[34] Einstein, later in his career, however expressed dissatisfaction and criticism of empiricist attitudes on a range of issues.[35]

His short message to the 1952 Michelson centenary, along with its citations of the Fizeau experiment and of aberration, implied the denial of a strong role for the Michelson-Morley experiment. It pointedly ended with: "There is, of course, no logical way leading to the establishment of a theory but only groping constructive attempts controlled by careful consideration of factual knowledge." This strongly suggests that Einstein's intended audience were overambitious empiricists. As he wrote in his intellectual autobiography of 1949, the "positivistic philosophical attitude" suffered from a prejudice: "the prejudice consists in the faith that facts by themselves can and should yield scientific knowledge without free conceptual construction."[36]

---

[29] Ludwik Silberstein to Einstein, 4 September 1921 (CPAE 12, Doc. 229); Einstein to Silberstein, 4 October 1921 (CPAE 12, Doc. 254).

[30] On the bias in Einstein's historical accounts, see (van Dongen 2002), (Janssen and Renn 2007).

[31] On this shift, see (Holton 1968/1995), (Norton 2000), (van Dongen 2002), (van Dongen 2004); for Einstein's self-identification as a "believing rationalist" that "seeks the only trustworthy source of truth in mathematical simplicity", see his letter to Cornelius Lanczos of 24 January 1938, cited on p. 259 in (Holton 1968/1995).

[32] See for instance (Hentschel 1992) and (van Dongen 2007), in particular its pp. 116-118.

[33] "But I was also guided by the result of the Fizeau experiment and the phenomenon of aberration." (Shankland 1964), p. 35; see also (Holton 1969/1995), p. 303, and EA 1 168; for an interpretation of these comments, see (Norton 2004), pp. 82-92.

[34] See (Holton, 1969/1995), in particular pp. 293-298.

[35] See e.g. his "Autobiographisches" (Einstein 1949a), in particular pp. 49, 80; "Reply to criticisms" (Einstein 1949b), p. 678.

The Michelson-Morley experiment is only one among a mêlée of factors that put Einstein on the path to the special theory, including, e.g. the thought-experiment of moving coil and magnet. His Chicago lecture does not warrant a return to the old lore as promulgated by Millikan. Yet, the contrast it exhibits with the older Einstein's pronouncements on the role of the experiment casts doubt on the accuracy of his late personal recollections, and draws attention to their specific philosophical context.

*Acknowledgments*: I am most grateful to Diana Kormos Buchwald for the suggestion to write this article and for very useful commentary to an earlier version. I also wish to thank Dennis Dieks, A.J. Kox, John Norton and Tilman Sauer for sharing their insight with me. Finally, I wish to thank Brenda Shorkend of the Einstein Papers Project and Harry Miller of the Wisconsin Historical Society for unearthing Einstein's Chicago lectures.

---

[36] (Einstein 1949a), on p. 49.